\documentclass[prb,twocolumn,showpacs,floatfix,preprintnumbers,amsmath,amssymb,superscriptaddress]{revtex4}

\usepackage{graphicx}
\usepackage{amsmath}
\usepackage{amssymb}
\usepackage{color}
\usepackage{dcolumn}
\usepackage{epsfig}
\usepackage{bm}
\usepackage[urlcolor=blue]{hyperref}
\hypersetup{backref, colorlinks=true, linkcolor=blue, citecolor=blue}

\begin{document}

\preprint{}

\title{Nodeless superconductivity in the presence of spin-density wave in pnictide superconductors: The case of BaFe$_{2-x}$Ni$_{x}$As$_{2}$}

\author{Mahmoud Abdel-Hafiez}
\affiliation{Center for High Pressure Science and Technology Advanced Research, Shanghai, 201203, China}

\author{Yuanyuan Zhang}
\affiliation{Key Laboratory of Polar Materials and Devices, East China Normal University, Shanghai 200241, China}

\author{Zheng He}
\affiliation{Center for High Pressure Science and Technology Advanced Research, Shanghai, 201203, China}
\affiliation{State Key Laboratory of Surface Physics and Department of Physics, Fudan University, Shanghai 200433, China}

\author{Jun Zhao}
\affiliation{State Key Laboratory of Surface Physics and Department of Physics, Fudan University, Shanghai 200433, China}
\affiliation{Collaborative Innovation Center of Advanced Microstructures, Nanjing 210093, China}

\author{Christoph Bergmann}
\affiliation{Max Planck Institute for Chemical Physics of Solids, 01187 Dresden, Germany}

\author{Cornelius Krellner}
\affiliation{Institute of Physics, Goethe University Frankfurt, 60438 Frankfurt/M, Germany}

\author{Chun-Gang Duan}
\affiliation{Key Laboratory of Polar Materials and Devices, East China Normal University, Shanghai 200241, China}

\author{Xingye Lu}
\affiliation{Beijing National Laboratory for Condensed Matter Physics, Institute of Physics, Chinese Academy of Sciences, Beijing 100190, China}

\author{Huiqian Luo}
\affiliation{Beijing National Laboratory for Condensed Matter Physics, Institute of Physics, Chinese Academy of Sciences, Beijing 100190, China}

\author{Pengcheng Dai}
\affiliation{Beijing National Laboratory for Condensed Matter Physics, Institute of Physics, Chinese Academy of Sciences, Beijing 100190, China}
\affiliation{Department of Physics and Astronomy, Rice University, Houston, Texas 77005, USA}

\author{Xiao-Jia Chen}
\email{xjchen@hpstar.ac.cn}
\affiliation{Center for High Pressure Science and Technology Advanced Research, Shanghai, 201203, China}

\date{\today}

\begin{abstract}

The characteristics of Fe-based superconductors are manifested in their electronic, magnetic properties, and pairing symmetry of the Cooper pair, but the latter remain to be explored. Usually in these materials, superconductivity coexists and competes with magnetic order, giving unconventional pairing mechanisms. We report on the results of the bulk magnetization measurements in the superconducting state and the low-temperature specific heat down to 0.4 K for BaFe$_{2-x}$Ni$_{x}$As$_{2}$ single crystals. The {electronic} specific heat displays a pronounced anomaly at the superconducting transition temperature and a small residual part {at low temperatures in the superconducting state}. The normal-state Sommerfeld coefficient increases with Ni doping for $x$ = 0.092, 0.096, and 0.10, which illustrates the competition between magnetism and superconductivity. Our analysis of the temperature dependence of the superconducting-state specific heat and the London penetration depth provides strong evidence for a two-band $s$-wave order parameter. Further, the data of the London penetration depth calculated from the lower critical field follow an exponential temperature dependence, characteristic of a fully gapped superconductor. These observations clearly show that the superconducting gap in the nearly optimally doped compounds is nodeless.

\end{abstract}


\maketitle

\section{Introduction}

One of the major themes in the physics of condensed matter is unconventional superconductivity in Fe-based materials~\cite{Kam,Ding,d,pag}. These materials have multiple Fermi pockets with electronlike and holelike dispersion of carriers and both hole and electron Fermi pockets show a low carrier density~\cite{RAE}. Superconductivity appears at the border of the antiferromagnetic (AF) regime, which may have a significant impact on the pairing mechanism~\cite{Park}. However, the exact picture of the interplay between superconductivity and magnetism remains elusive~\cite{Ding}. Although other scenarios involving orbital fluctuations are possible, it has generally been believed that spin fluctuations play an important role and act as the mediating bosons for electron pairing and superconductivity~\cite{Hirschfeld}. Despite great successes in studying these materials, there are still unresolved issues, particularly the symmetry and structure of the order parameter, and doping evolution of the superconducting (SC) gap, which should provide an understanding of the pairing mechanism of these systems~\cite{pag,Hirschfeld,Thomale}. It has been well characterized that both cuprates and conventional phonon-mediated superconductors are characterized by distinct $d$-wave and $s$-wave pairing symmetries with nodal and nodeless gap distributions, respectively. There is no general consensus on the nature of pairing in iron-based superconductors leaving the perspectives ranging from $S^{++}$ wave, to $S^{\pm}$, and to $d$ wave~\cite{PRB,US,PP,STJ,Dong,Tort,FH,Ding1,Ding2,Ding3,Tan,JHu,Kre,YZh,M1,M2}.  In addition, from  $^{59}$Co and $^{75}$As nuclear magnetic resonance measurements, the spin triplet order parameter was ruled out in BaFe$_{1.8}$Co$_{0.2}$As$_{2}$~\cite{NMR}. It turns out that the SC gap distributions are  vary with different systems and unusually are sensitive to the sample quality. For systems with both hole and electron Fermi surfaces, such as optimally doped Ba$_{1-x}$K$_{x}$Fe$_{2}$As$_{2}$~\cite{PP}, Ba(Fe$_{1-x}$Co$_{x}$)$_{2}$As$_{2}$~\cite{FH}, NaFe$_{1-x}$Co$_{x}$As~\cite{Tan}, and Fe(Se, Te)~\cite{JHu}, the gaps measured by low temperatures specific-heat fit well to the predictions of two nodeless SC gap. The temperature dependence of the lower critical field in LiFeAs~\cite{L1}, Ba$_{0.6}$K$_{0.4}$Fe$_{2}$As$_{2}$~\cite{CR}, FeSe~\cite{V12}, and Ca$_{0.32}$Na$_{0.68}$Fe$_{2}$As$_{2}$~\cite{PRB} has supported the existence of two s-wave-like gaps. The possibility of nodes along the $c$ axis in the superconducting gap has been reported in NdFeAsO$_{0.82}$F$_{0.18}$ and LaFePO, where the magnetic penetration depth exhibited a nearly linear temperature dependence~\cite{XLW,Hi}.

We begin with listing several facts about BaFe$_{2-x}$Ni$_{x}$As$_{2}$. (i) In the first, it appears as an ideal candidate to study the fundamental properties of superconductivity due to the availability of high-quality single crystals with rather large dimensions~\cite{S1}. (ii) In the undoped state, BaFe$_{2}$As$_{2}$ shows a combined spin-density wave (SDW) and structural transition near $T_{N}$ = $T_{s}$ = 138\,K. The pristine compound is characterized by a bad metallic behavior with a coherent Drude component and doping with Ni and P transforms a bad metal to a good metal, while the system remains a bad metal with K-doping~\cite{Nat}. (iii) The N\'{e}el temperature of the electron-doped iron pnictides decreases gradually with increasing electron-doping level, and the AF phase appears to coexist with the SC phase~\cite{Ni,Di,Di1}. However, a neutron-scattering study reveals an avoided quantum critical point, which is expected to influence the properties of both the normal and SC states strongly~\cite{Di1}. This raises a critical question concerning the role of quantum criticality~\cite{Abra} and the coexistence of magnetism and superconductivity to the SC pairing structure~\cite{Di}. (iv) Furthermore, a recent neutron-scattering measurement has revealed that the low-energy spin excitations in BaFe$_{2-x}$Ni$_{x}$As$_{2}$ change from fourfold symmetric to twofold symmetric at temperatures corresponding to the onset of the in-plane resistivity anisotropy. In the overdoped compounds both resistivity and spin excitation anisotropies are vanished. Therefore, they are likely intimately connected~\cite{d}. (vi) The London penetration depth $\lambda$ measurements suggest that the competition between superconductivity and magnetic/nematic order in hole-doped compounds is weaker than in electron-doped compounds~\cite{kim}. In this context, it is important to understand the doping, field, and temperature dependence of AF spin correlations. Studying the symmetry and structure of the order parameter is a key not only to understand all these interesting features but also to address unsettled issues in BaFe$_{2-x}$Ni$_{x}$As$_{2}$.

Low-temperature specific heat $C_{P}$ and the London penetration depth $\lambda$ are two powerful techniques for probing the gap structure of bulk superconductors. Both measurements probe bulk SC properties. $\lambda$ is a fundamental parameter which detects the pairing symmetry and the $T$-dependence of $\lambda$ can determine gap function. Since $C_{P}$ is directly related to the quasiparticle density of states, its temperature dependence reflects the nature of the SC state such as gap symmetry, the presence of multigaps, and coupling strength between electrons and phonons. In addition, it is less affected by vortex pinning. An exponential vanishing of the specific heat at low temperature in conventional $s$-wave superconductors is caused by the finite gap in the quasiparticle spectrum. This is due to the quasiparticle thermal fluctuations go exponentially to zero as $T\rightarrow$ 0. For SC gap with gap nodes, electronic excitations  are possible even at very low temperatures~\cite{M2}. In general, specific heat comprises of two parts: the electronic $C_{el}$ and the phononic contribution $C_{ph}$. Information about the pairing symmetry is contained in the $C_{el}$, which is proportional to the density of states (DOS) at the Fermi energy. Exploring the symmetry and structure of the order parameter, and the evolution of the SC gap with Ni doping in BaFe$_{2-x}$Ni$_{x}$As$_{2}$ system based on the mentioned two bulk detection techniques is thus highly desired. It should be mentioning that we have estimated the $C_{ph}$ from BaFe$_{1.75}$Ni$_{0.25}$As$_{2}$. This sample is not superconductor throughout the temperature range as evident in Fig.\,2(e) where the $C/T$ exhibits a monotonous increase against the temperature. The fact that the low temperature specific heat data for the investigated samples [the inset of Fig.\,2(e)] {exhibit} a linear behavior at low temperatures without any upturn indicates the absence of {Schottky-like} contributions in our samples. Furthermore, at $T >$ T$_{c}$ the specific heat data of the SC and non-SC samples are comparable, confirming similar phonon contributions to the specific heat of SC and non-SC samples. Therefore, magnetic contribution to specific heat will be negligible and the specific heat can be assumed to have contribution from the electronic and lattice part only.

In this work, we use magnetization and low temperature specific heat measurements on BaFe$_{2-x}$Ni$_{x}$As$_{2}$ to study the interplay between magnetism and superconductivity with the emphasis of nature of the SC pairing symmetry by focusing on materials near optimal doping [Fig.\,1(a)]. Based on the comprehensive low-$T$ measurements, we provide evidence for nodeless superconductivity in the doping range of $x$ = 0.092, 0.096, and 0.10. The temperature-dependence of $\lambda _{ab}(T)$ calculated from the lower critical field and the $C_{el}$ can be well described by using a two-band model with $s$-wave-like gaps. Reliable values of the normal-state Sommerfeld coefficients are obtained for the studied materials, which increases with Ni doping, illustrating the strong competition between magnetism and superconductivity.

\begin{figure*}[tbp]
\includegraphics[width=43pc,clip]{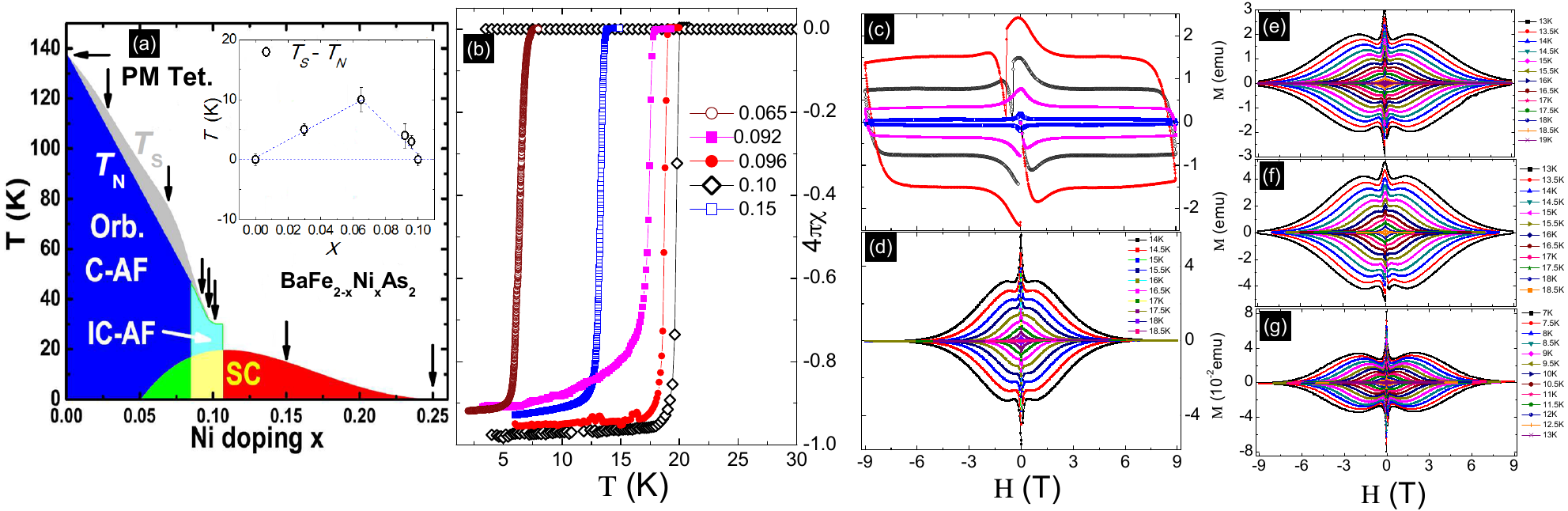}
\caption{\label{fig:wide} (a) The electronic phase diagram of BaFe$_{2-x}$Ni$_{x}$As$_{2}$ obtained from magnetic and specific heat data, showing the suppression of the magnetic ($T_{N}$) and structural ($T_{S}$) phase transitions with increasing Ni concentration and the appearance of the SC transitions. The arrows indicate eight doping levels studied in this work. The PM Tet, PM Orb, C-AF, and IC-AF are paramagnetic tetragonal, paramagnetic orthorhombic, commensurate AF orthorhombic, and incommensurate AF orthorhombic phases, respectively. The inset illustrates the electron-doping dependence of $T_{S} - T_{N}$. (b) shows the  temperature dependence of the magnetic susceptibility in an external field of 10\,Oe applied along the $c$ axis. The susceptibility has been deduced from the dc magnetization measured by following ZFC and FC protocols of BaFe$_{2-x}$Ni$_{x}$As$_{2}$ single crystals. (c) presents the isothermal magnetization $M$ vs. $H$ loops measured at 2\,K up to 9\,T for $H \parallel c$ for $x$ = 0.092, 0.096, 0.10, and 0.15. (d) 14-18.5\,K for each 0.5\,K, (e) 13-18.5\,K for each 0.5\,K, (f) 13-19\,K for each 0.5\,K, and (g) 7-13\,K for each 0.5\,K plots at high temperatures exhibit a pronounced second peak for $x$ = 0.092, 0.096, 0.10, and 0.15 respectively.}
\end{figure*}

\section{Experimental}

BaFe$_{2-x}$Ni$_{x}$As$_{2}$ ($x$ = 0, 0.03, 0.065, 0.092, 0.096, 0.10, 0.15, and 0.25) single crystals were grown by the FeAs self-flux method~\cite{S1}. The actual Ni level was determined to be 80\% of the nominal level $x$ through the inductively coupled plasma analysis of the as-grown single crystals. Magnetization measurements were performed by using a Quantum Design SC quantum interference magnetometer. The low-$T$ specific heat down to 0.4\,K was measured in its Physical Property Measurement System with the adiabatic thermal relaxation technique along $H \parallel c$ up to $H$ = 9\,T.

\section{Results and discussion}

The arrows in Fig.\,1(a) indicate eight doping levels presented in this work. These include $x$ = 0 (parent compound shows a $T_{S}$($T_{N}$) = 137(2)\,K), $x$ = 0.03 and 0.065 (lightly electron-doped non SC and SC samples with $T_{S}$/$T_{N}$ = 110/104 and 82/70\,K) respectively, $x$ = 0.092, 0.096, and 0.10 (nearly optimal doping SC samples with static incommensurate short-range order), $x$  = 0.15 (overdoped superconducting sample without AF order coexisting with superconductivity) and $x$ = 0.25 (heavily overdoped non SC sample). As for BaFe$_{2-x}$Co$_{x}$As$_{2}$ system~\cite{S3c}, it has been shown that near optimal superconductivity (see Fig.\,1(a)), the commensurate static AF order changes into transversely incommensurate short-range AF order that coexists and competes with superconductivity~\cite{S3}. Similar to the case of BaFe$_{2-x}$Co$_{x}$As$_{2}$ and CaFe$_{2-x}$Co$_{x}$As$_{2}$, the underdoped region exhibits a splitting of the structural and magnetic phase transitions. The inset of Fig.\,1(a) shows the electron-doping dependence of $T_{S} - T_{N}$. Figure 1(b) shows the magnetic susceptibility measured with the zero field cooling (ZFC) and field cooling (FC) in an external field of 10\,Oe applied along the $c$ axis. The FC and ZFC data prove a sharp diamagnetic signal. Beyond, the SC volume fraction is close to 1, thus confirming bulk superconductivity and the high quality of BaFe$_{2-x}$Ni$_{x}$As$_{2}$ single crystals. The $T_{c}$ has been determined from the onset diamagnetic transition temperature between ZFC and FC to be around $\sim$ 7.7, 18.5, 19.0, 20.0, and 13.9\,K for $x$ = 0.065, 0.092, 0.096, 0.10, and 0.15 respectively. The clear irreversibility between FC and ZFC measurements is the consequence of a strong vortex trapping mechanism, either by surface barriers or bulk pinning.

\begin{figure*}[tbp]
\includegraphics[width=43pc,clip]{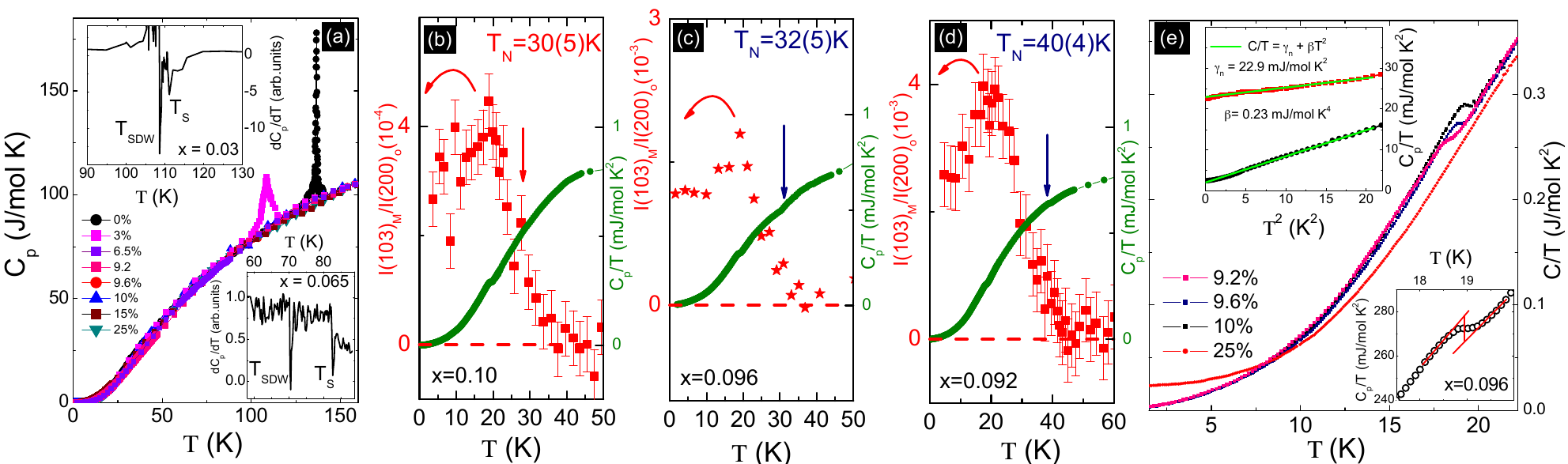}
\caption{\label{fig:wide} (a) Temperature dependence of the specific heat of BaFe$_{2-x}$Ni$_{x}$As$_{2}$ ($x$ = 0, 0.03, 0.065, 0.092, 0.096, 0.10, 0.15, and 0.25) measured in zero magnetic field. The insets show the derivative of specific heat for the crystal with $x$ = 0.03 (upper inset) and  $x$ = 0.065 (lower inset), where structural and SDW transitions can be clearly recognized from the dips. (b) $x$ = 0.10, (c) $x$ = 0.096, and (d) $x$ = 0.092 indicated the neutron data counting time (30 min=point on HB-1A taken from Ref.~[\onlinecite{Di1}]) and the specific heat. The $T_{N}$ is marked by an arrow. (e) The temperature dependence of the specific heat $C/T$ of samples with $x$ = 0.092, 0.096, 0.10, and 0.25 down to $T$ = 400\,mK. The upper inset shows the low-temperature specific heat of two samples with $x$ = 0.10 and 0.25. The straight lines represent linear fits to $C_{p} = \gamma T + \beta T^{3}$. The lower inset presents the enlarged  $C_{p}/T$ vs. $T$ plot near the SC transition for $x$ = 0.096. The lines show how $C_{p}/T_{c}$ and $T_{c}$ are estimated.}
\end{figure*}

Figure\,1(c) presents the field dependence of the isothermal magnetization $M$ at 2\,K up to 9\,T for $H \parallel c$ for $x$ = 0.092, 0.096, 0.10, and 0.15. At $T$ = 2\,K for $x$ = 0.096 and 0.10, the $M(H)$ exhibits irregular jumps close to $H$ = 0 similarly to LiFeAs, Ba$_{0.65}$Na$_{0.35}$Fe$_2$As$_2$, and Ca$_{0.32}$Na$_{0.68}$Fe$_2$As$_2$ superconductors~\cite{V11a,V11,PRB}. Figures\,1(d-g) present the field dependence of the isothermal magnetization $M$ at various temperatures very close to $T_{c}$ up to 9\,T for $x$ = 0.092, 0.096, 0.10 and 0.15 respectively. In addition, the SC $M(H)$ exhibits no magnetic background. This indicates that our investigated samples contain negligible magnetic impurities. The width of the magnetic loops decrease while increasing the applied field. However, at higher temperatures the width of the loops initially decreases showing a minimum at the $H_{m}$ field and then increases again. Further, the $M(H)$ loops demonstrate another pronounced peak or so-called second peak. The second peak effect has been studied extensively and its origin may be attributed to various mechanisms. It has been well established that the second peak effect is strongly influenced by the oxygen deficiency in cuprates~\cite{P1,P2}. In the case of  Fe-based superconductors, the local magnetic moments may form the small size normal cores, and may be a possible reason of the second peak effect~\cite{V3}. However, the real pinning mechanism needs further investigation. The position of the second peak shifts to higher fields while decreasing temperature, eventually beyond the available field range. This can explain the nonvisibility of a second peak at low temperatures in the Figs.\,1(d-g). The $M(H)$ loops show irreversibility in magnetization, which vanishes above a characteristic field $H_{irr}$ [Figs.\,1(d-g)]. It is noteworthy that the first vortex penetration field may not reflect the true $H_{c1}(T)$ because of Bean-Livingston surface barrier. The fact that the hysteresis loops for $x$ = 0.092, 0.096, 0.10, and 0.15 are symmetric around $M = 0$, pointing to relatively no surface barriers and implying that the bulk pinning plays a dominant role in our investigated compounds. In contrast to that, if surface barriers were predominant, the first vortex entrance can occur at much higher field ($\approx H_{c}$). This is a very important point in order to obtain reliable estimations of the thermodynamic lower critical field (see below).

Specific heat provides a probe for the symmetry and structure of the SC order parameter. Figure\,2(a) summarizes the temperature dependence of the zero-field specific-heat data at various Ni-doping levels in the BaFe$_{2-x}$Ni$_{x}$As$_{2}$ series plotted as $C_{p}$ vs. $T$. The data of the parent-compound ($x$ = 0) shows a very sharp first-order structural transition coinciding with the SDW transition at 136\,K (upon heating) and with a transition width of about 3\,K. Because of the narrowness of the transition, a temperature rise of only 0.5~\% was used for each measurement in the vicinity of the transition of all measurements. Upon Ni-doping, the sharp first-order structural/magnetic anomaly of the parent compound gradually broadens, shifts and splits to lower temperatures and is considerably reduced in magnitude. For $x$ = 0.03 and 0.065, the combined structural/magnetic anomaly of the pristine compound actually splits into two distinct anomalies at 110, 104 and 86, 74\,K, respectively. The error in the determination of the $T_{S}$ and $T_{N}$ transition temperatures can be estimated at around 2\,K if we take into account that the peak in the first derivative of the specific heat is relatively sharp [see upper and lower insets of Fig.\,2(a)]. Then, the transition is shifted to 40(4), 30(5), and 32(5)\,K for $x$ = 0.092, 0.096, and 0.10, respectively [see Fig.\,2(b-d)]. These data are in line with the recent high-resolution x-ray and neutron scattering data as discussed in Ref.~[\onlinecite{Di1}]. Recent neutron scattering data on $x$ = 0.10 sample reveal a weak static AF order with magnetic scattering 5 times smaller than that of $x$ = 0.096. In spite of the small moments of $x$ = 0.10, the temperature dependence of the magnetic order parameters for both samples indicates that their AF temperatures are essentially unchanged at $T_{N} \pm 5$\,K~[\onlinecite{Di1}].

Figure\,2(e) shows the temperature dependence of the specific heat of the samples with $x$ = 0.092, 0.096, 0.10, and 0.25 down to 0.4\,mK. An entropy conserving construction has been used to determine the SC transition temperature from the specific heat data. For $x$ = 0.092, 0.096, and 0.10 a clear anomaly at 18.4, 18.9, 20\,K respectively indicates the onset of bulk superconductivity. The sample, with $x$ = 0.25, remains in the normal state. The fact that the low temperature specific heat data {exhibit} a linear behavior at low temperatures without any upturn indicates the absence of {Schottky-like} contributions in our investigated samples [see upper inset of Fig.\,2(e)]. It is important to note that it is impossible to obtain the lattice background by fitting the specific heat of the SC samples to an odd-power polynomial above $T_{c}$ due to the electronic term of the total signal of the specific heat data. As demonstrated below, a more reliable phonon term can be estimated from the data of the $x$ = 0.25 sample, whose low-temperature specific heat follows precisely the Debye law between 0.4 and 4.5\,K, with $\gamma _{n}$ = 22.9\,mJ/mol K$^{2}$ and $\beta$ = 0.23\,mJ/mol K$^{4}$.

Further experimental investigations on the structure and magnitude of the SC gaps in BaFe$_{2-x}$Ni$_{x}$As$_{2}$ by means of bulk specific heat data are of great interest. In order to determine the specific heat related to the SC phase transition we need to estimate the $C_\mathrm{ph}$ and $C_\mathrm{el}$ contributions to $C_p$ in the normal state. In order to determine the phononic contribution to the specific heat for $x$ = 0.25, the following relation is used: $C^{x=0.25}_{Ph} = C^{x=0.25}_{tot} - C^{x=0.25}_{el} $, where $C_{el}^{x=0.25}$ is {just} $\gamma T$. {The same shape of the phononic heat capacity in the SC samples and overdoped sample is assumed. Therefore, the specific heat of the SC samples can be represented by:
\begin{equation}\label{eq1}
C_{el}^{SC}/T = C_{tot}^{SC}/T - \emph{g}. C_{ph}^{x=0.25}/T,
\end{equation}
which allows us to calculate the $C_\textup{el}$} of the SC samples. The small deviation of the scaling factor \emph{g} from unity,  plausibly related to experimental uncertainties, demonstrates that the above procedure represents a very good method to determine the phonon background. The value of $g$ was determined from the requirement of equality between the normal and SC state entropies at $T_\textup{c}$, {that is} $\int_{0}^{T_\textup{c}}\left(C_\textup{el}/T\right)dT = \gamma_\textup{n} T_\textup{c}$, where $\gamma_\textup{n}$ is the normal state electronic specific heat coefficient. We started with $g= 1$, but we found that the entropy conservation criterion is satisfied with {$g = 0.95$}. Physically, this indicates that the substitution of Fe by Ni does not substantially affect the lattice properties.

\begin{figure}[tbp]
\includegraphics[width=18pc,clip]{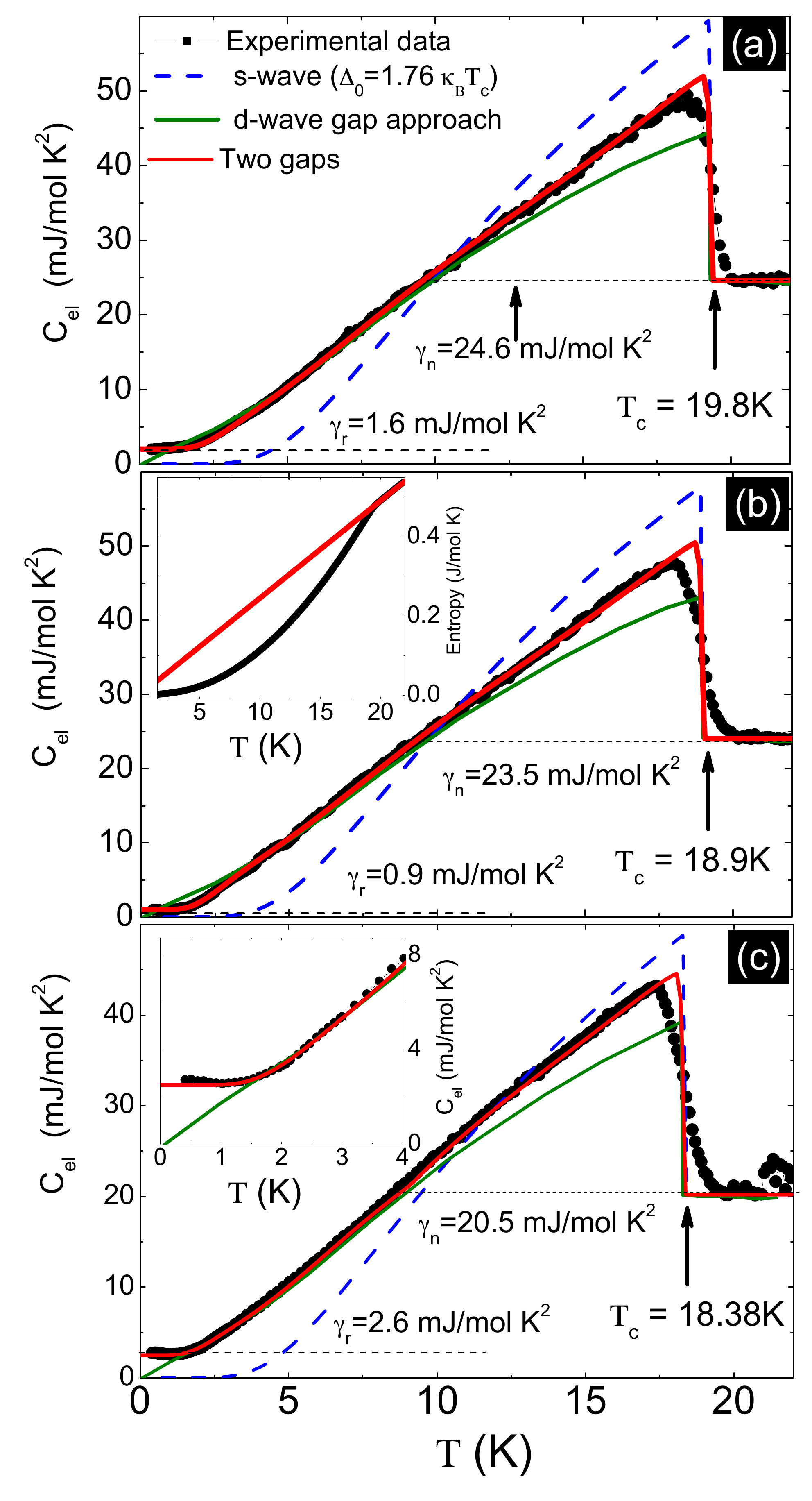}
\caption{The electronic specific heat C$_{el}$/T as a function of temperature for BaFe$_{2-x}$Ni$_{x}$As$_{2}$ [$x$ = 0.1 (a), 0.096 (b), and 0.092 (c)]. The inset in (b) presents the entropy in the normal and superconducting state as a function of $T$. The inset in (c) shows the low-$T$ data on a larger scale. $\gamma$$_{n}$ represents the normal-state electronic coefficient of the specific heat and $\gamma_\textup{r}$ is the residual electronic specific heat. The dashed lines represent the theoretical curves based on single-band weak-coupling BCS theory, while the solid lines illustrate the $d$-wave approximation. The solid red lines indicate the curves of the two $s$-wave gap model.}
\label{PD}
\end{figure}

Figure\,3 shows the temperature dependence of the electronic contribution to the specific heat in the zero field determined by subtracting $C_\mathrm{ph}$ for $x$ = 0.10 [Fig.\,3(a)], 0.096 [Fig.\,3(b)] and 0.092 [Fig.\,3(c)]. The entropy conservation required for a second-order phase transition is fulfilled as shown in the inset of Fig.\,3(b). This check warrants the thermodynamic consistency for both, the measured data and the determination of $C_\mathrm{el}$. It is obvious from Fig.\,3 that the SC transition at $T_\textup{c}$ is well pronounced showing a sharp jump in $C_\mathrm{el}$ at $T_c$. The jump height of the specific heat at $T_c$ is found to be $\Delta C_{el}/T_{c}$ $\approx$ 23(0.5), 24.8(2), and 25.1(1)~mJ/mol K$^2$ for $x$ = 0.092, 0.096, and 0.10, respectively. Generally, the specific heat jumps at $T_c$ obtained for these materials scale relatively well with its $T_c$ in light of the recent careful results for the pnictide superconductors~\cite{pag,SLB2} in which the universal curve $\Delta C_{p}/T_{c} \varpropto T^{3}$ is explained. Furthermore, it has been well reported that the jump of the specific heat $\Delta C/T_{c}$ varies with $T_{c}$, and has a peak near optimal doping and decreases at smaller and larger doping. This is a direct manifestation of the coexistence between antiferromagnetism and SC order parameters~\cite{Vav}. From our determined $\gamma_\textup{n}$ = 20.5, 23.5, and 24.6\,mJ/mol K$^{2}$ for $x$ = 0.092, 0.096, and 0.10 respectively, we find $\Delta C_\textup{el}/ \gamma_\textup{n} T_\textup{c}$ = 1.1, 1.06, and 1.04 for $x$ = 0.092, 0.096, and 0.10 respectively. These values are smaller than the prediction of the weak coupling BCS theory ($\Delta C_\textup{el}/ \gamma_\textup{n} T_\textup{c} = 1.43$). Taking {into} account the fact that the SC transition is relatively sharp in our SC samples, a distribution in $T_\textup{c}$ or the presence of impurity phases {cannot} explain the reduced value of the specific heat jump. In addition, $\gamma_\textup{n}$ increases with Ni doping, illustrating the competition between magnetism and superconductivity.

We believe, however, that the presence of multiple SC gaps {may} reduce the universal parameter, as evidenced in other 122 Fe-based superconductors~\cite{Hardy2010}. It has been also well reported that the reduced jump in the specific heat $\Delta C_{p}/T_{c}$ compared to that of a single-band $s$-wave superconductor might be related to a pronounced multiband character with rather different partial densities of states and gap values~\cite{M2}. Note that $C_\mathrm{el}/T$ almost saturates at low temperature; however, it does not extrapolate to zero, yielding a residual electronic specific-heat value $\gamma_\textup{r}$ = 2.6, 0.9, and 1.6\,mJ/mol K$^{2}$ for $x$ = 0.092, 0.096, and 0.10, respectively. The finite value of $\gamma_\textup{r}$ indicates a finite electronic density of states at low energy, even in zero applied field. We mention that the presence of a finite $\gamma_\textup{r}$ is common in both electron- and hole-doped 122 crystals and that the value of $\gamma_\textup{r}$ is remarkably low, showing the good quality of our investigated single crystals. However, the origin of this residual term is still unclear. It may be because of an incomplete transition to the SC state or because of broken pairs caused by disorder or impurities in unconventional superconductors, and/or spin-glass behavior. On the other hand, previous specific heat measurements on optimally doped YBa$_{2}$Cu$_{3}$O$_{7-\delta}$ exhibit such a $\gamma_\textup{r}$ term. For instance, even the best YBa$_{2}$Cu$_{3}$O$_{6.56}$ samples present $\gamma_\textup{r}$ $\approx$ 1.85 mJ/mol K$^{2}$~[\onlinecite{Cup}]. It has been proposed that this $\gamma_\textup{r}$ term originates from a disorder-generated finite density of quasiparticle states near the $d$-wave nodes. It is worth to mention that $\gamma_\textup{r}$ in our SC samples reaches 12.6, 3.8, and 6.5\% of the normal state Sommerfeld coefficient $\gamma_\textup{n}$ for $x$ = 0.092, 0.096, and 0.10, respectively. A similar observation of ($\gamma_\textup{r}$/$\gamma_\textup{n}$ $\approx$ 5.7\%- 24\%) \cite{Pramanik2011,Hardy2010} was also reported in iron pnictide superconductors.

The almost linear temperature dependence of $C_\mathrm{el}/T$ of the SC samples indicates that the specific heat data cannot be described by a single BCS gap. In order to illustrate this we show a theoretical BCS curve with $\Delta = 1.764\, k_{\mathrm{B}}T_\mathrm{c} = 2.23$~meV in Fig.\,3. One can see that systematic deviations from the data are observed in the whole temperature range below $T_\mathrm{c}$. Since a single gap cannot describe the data, we applied a $d$-wave calculation and a phenomenological two-gap model developed for the specific heat of MgB$_2$~\cite{Bouquet2001} as in Eq.\,(2) and Eq.\,(3). For the $d$-wave approximation we used $\Delta = \Delta_{0}\cos(2\theta)$. In the case of a two-band model, the thermodynamic properties are obtained as the sum of the contributions from the individual bands, $i.e.$, $\alpha_{1} = \Delta_{1}/k_{B}T_{c}$ and $\alpha_{2} = \Delta_{2}/k_{B}T_{c}$
\begin{equation}\label{eq5}
    \frac{S}{\gamma_{n}T_{c}}=-\frac{6\Delta_{0}}{\pi^{2}k_{B}T_{c}}\int_{0}^{\infty}[f\ln f+ (1-f)\ln (1-f)]dy,
\end{equation}
\begin{equation}
\label{eq5}
 \frac{S}{\gamma_{n}T_{c}}= t\frac{d( \frac{C}{\gamma_{n}T_{c}})}{dt},
\end{equation}
where $t = T/T_{c}$, $f = [ \exp( \beta E + 1 )]^{-1}$, $\beta$ = ($k_\textup{B}T)^{-1}$, and the energy of the quasiparticles is given by $E$ = $[\epsilon^{2} + \Delta^{2}(t)]^{0.5}$ with $\epsilon$ being the energy of the normal electrons relative to the Fermi {level}. The integration variable is $y =\epsilon/\Delta_0$. In Eq.\,(2), the scaled gap $\alpha = \Delta_0/k_BT$ is the only adjustable fitting parameter in the case of a single gap. At the same time, $\gamma _{i}/\gamma _{n}$ ($i$ = 1, 2), which measure the fraction of the total normal electron density of states, are introduced as adjustable parameters. This fitting is calculated as the sum of the contributions from two bands by assuming independent BCS temperature dependencies of the two SC gaps.

\begin{figure}[tbp]
\includegraphics[width=21pc,clip]{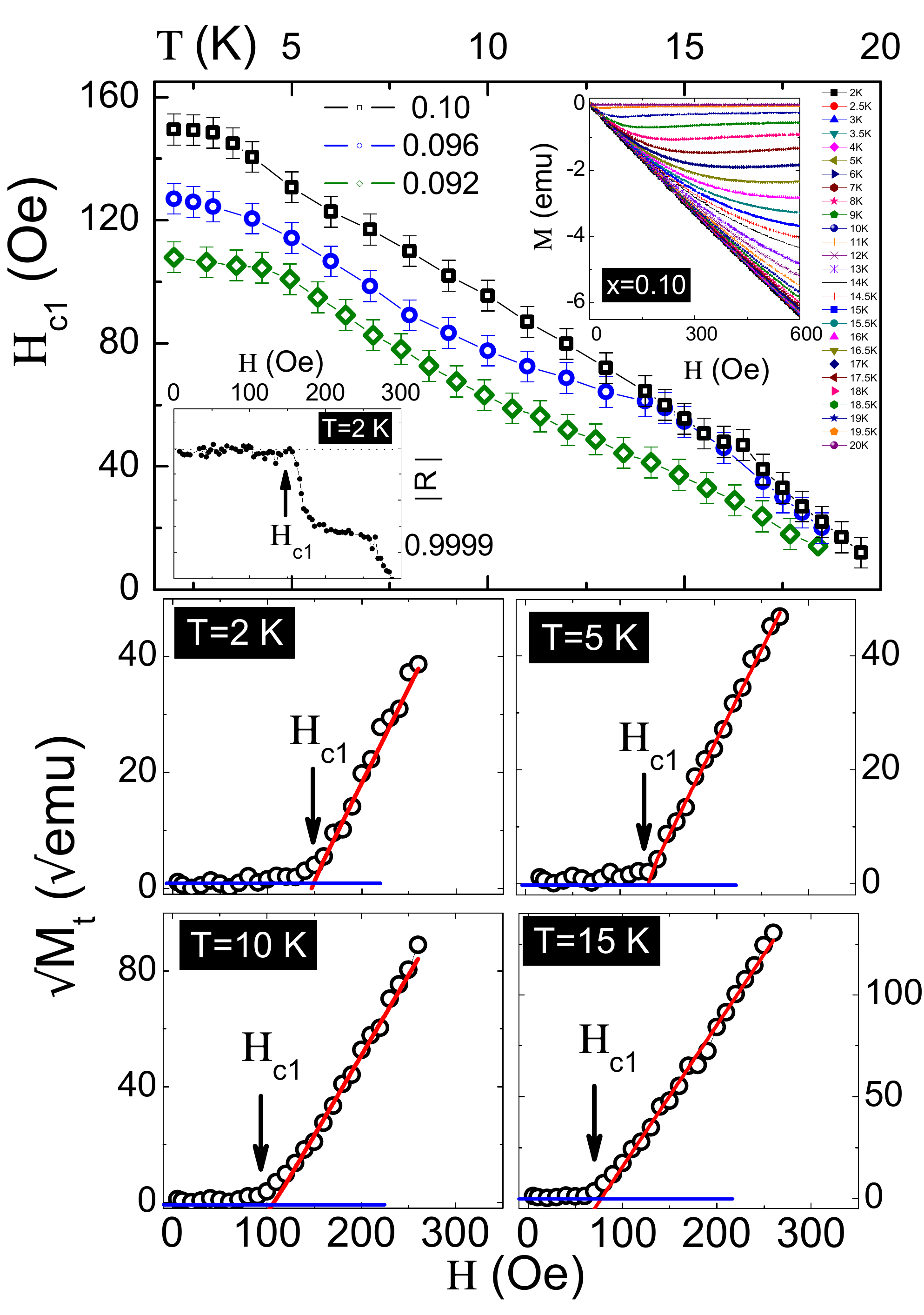}
\caption{The upper panel shows the phase diagram of $H_{\mathrm{c1}}$ vs.~the applied temperatures of BaFe$_{2-x}$Ni$_{x}$As$_{2}$ ($x$ = 0.092, 0.096, and 0.10) for the field applied parallel to the $c$ axis. The bars show the uncertainty of estimation by the deviating point of the regression fits. The error bar in the values of $H_{\mathrm{c1}}$ is about 5\,Oe of the investigated samples. The upper inset shows the field dependence of the superconducting initial part of the magnetization curves measured of BaFe$_{1.90}$Ni$_{0.10}$As$_{2}$ at various temperatures for $H \parallel c$. The lower inset depicts an example used to determine the $H _{c1}$ value using the regression factor, $R$, at $T$ = 2\,K. The lower panels present the field dependence of the typical plot of $\sqrt{M_{t}}$ vs $H$ at various temperatures for $x$ = 0.10. The solid lines are a linear fit to the high-field data of $\sqrt{M_{t}}$ vs. $H$. $H_{c1}$ values are determined by extrapolating the linear fit to $\sqrt{M_{t}}$ = 0.} \label{PD}
\end{figure}

The best description of the experimental data for each type of order parameter, $d$-wave and two-gaps $s$-wave can be seen in Fig.\,3. More obvious deviations exist in the case of the $d$-wave approach for the SC samples. This {clearly} indicates that the gap structure of our systems is more likely to be nodeless $s$-wave, which is reasonably well comparable with the penetration depth data (see below). The good description of the experimental data for the two-gaps $s$-wave model is obtained by using values of $\Delta _{1}(0)$ = 1.74, 1.8, and 1.85\,$k_{B}T_{c}$, $\Delta _{2}(0)$ = 0.68, 0.74, and 0.79\,$k_{B}T_{c}$ for $x$ = 0.092, 0.096, and 0.10, respectively. For the investigated systems, the large gap $\Delta_L$ has a higher value than the weak-coupling BCS (1.76$k_{B}T_{c}$) gap value, while the smaller one $\Delta_S$ has a value lower than the BCS one. This is consistent with the theoretical constraints that one gap must be larger than the BCS gap and one smaller in a weakly coupled two-band superconductor~\cite{TB}. Similar studies have been outlined in iron-based superconductors (Table I).

Next we discuss the temperature dependence of the lower critical field $H_{c1}$, the field at which vortices penetrate into the sample, in the SC-state, which is another independent test sensitive to the gap structure. However, determining the  $H_{c1}$ from magnetization measurements has never been an easy task. In order to determine the exact values of the $H_{c1}$ from the low-field $M$-$H$ curves measured at different temperatures, we have to detect the onset of the small deviation from the perfect diamagnetic signal. This is rather difficult and sometimes a debatable process. The most popular method to estimate $H_{c1}$ consists of detecting the transition from a Meissner-like linear $M(H)$ regime to a non-linear $M(H)$ response (see the upper inset of Fig.\,4 upper panel), once the vortices penetrate into the sample and build up a critical state. This transition is not abrupt therefore bearing a substantial error bar. These sort of measurements are obtained by tracking the virgin $M(H)$ curve at low fields at several temperatures, as shown in Fig.\,4 for $H \parallel c$ for $x$ = 0.092, 0.096, and 0.10. These $M$-$H$ curves show at low $H$ a linear dependence of magnetization on the field indicative of Meissner phase as well deviation from linearity at higher fields. We have adopted a rigorous procedure ($i.e.$ user-independent outcome) to determine the transition from linear to non-linear $M(H)$, which consists of calculating the regression coefficient $R$ of a linear fit to the data points collected between $0$ and $H$, as a function of $H$. Then, $H_{c1}$ is taken as the point where the function $R(H)$ departs from 1. The result of these calculations is illustrated in the lower inset of Fig.\,4 upper panel. Additionally, the temperature dependence of the first vortex penetration field has been experimentally obtained by measured the onset of the trapped flux moment $M_{t}$ as described in Refs.~[\onlinecite{Mosh,Hc1}]. In contrast to tracking the virgin $M(H)$ curves at low fields at several temperatures where a heavy data post-processing is needed now a careful measurement protocol needs to be followed with little data analysis. Indeed, the $H_{c1}$ values obtained from the onset of the $M_{t}$ are close to those obtained from the latter method.

Once the values of $H_{c1}$ have been experimentally determined, we need to correct them accounting for the demagnetization effects. Indeed, the deflection of field lines around the sample leads to a more pronounced Meissner slope given by $M/H_{a} = -1/(1-N)$, where $N$ is the demagnetization factor. Taking into account these effects, the absolute value of $H _{c1}$ can be estimated by using the relation proposed by Brandt~\cite{Bra}:
\begin{equation}
\label{eq2} q_{disk} = \frac{4}{3\pi}+\frac{2}{3\pi}{\tanh[1.27\frac{b}{a}\ln(1+\frac{a}{b})]},
\end{equation}
where $q \equiv (|M/H_{a}|-1)(b/a)$, and $a$ is the average of the dimensions perpendicular to the field of our investigated sample. For our samples we find $N$ $\approx$ 0.958(0.1), 0.95(0.12) and 0.94(0.085) for $x$ = 0.092, 0.096, and 0.10, respectively. The corrected values of $H _{c1}$ obtained by following the two methods described above, are illustrated in the main panel of Fig.\,4 for $H \parallel c$. In fact, the determination of $H_{c1}$ allows us to extract the magnetic penetration depth, a fundamental parameter characterizing the SC condensate which carries information about the underlying pairing mechanism. In the SC state, the temperature dependence of the penetration depth is a sensitive measure of low-energy quasiparticles, making it to a powerful tool for probing the SC gap~\cite{L1}.  In order to shed light on the pairing symmetry in our system, we estimated the penetration depth at low temperatures using the traditional Ginzburg-Landau theory, where $H_{c1}$ is given by: $\mu_{0}H_{c1}^{\parallel c} = (\phi$$_{0}/4\pi\lambda _{ab}^{2})\ln\kappa _{c}$, where $\phi$$_{0}$ is the magnetic-flux quantum $\phi$$_{0}$ = $h/e^{\ast}$ = 2.07$\times$10$^{-7}$Oe cm$^{2}$, $\kappa _{c}$ =$\lambda _{ab}$/$\xi _{ab}$ is the Ginzburg-Landau parameter~\cite{k1,k2}, which we obtained at $\lambda$(0) =214(15),  255(10), and 240(10)\,nm for $x$ = 0.092, 0.096, and 0.10, respectively.

\begin{figure}[tbp]
\includegraphics[width=21pc,clip]{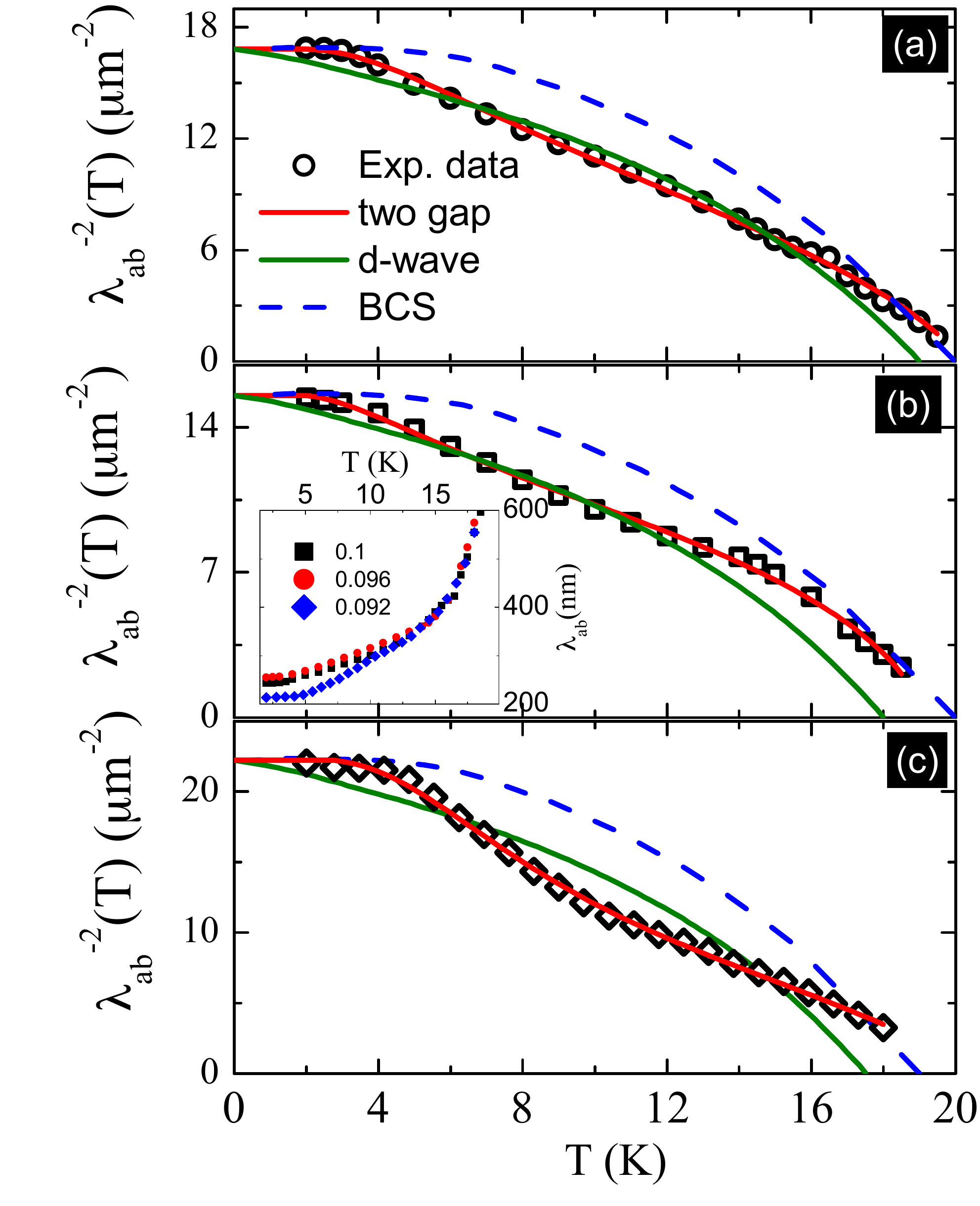}
\caption{The $T$-dependence of the $\lambda _{ab}(T)$ for BaFe$_{2-x}$Ni$_{x}$As$_{2}$ [$x$ = 0.10 (a), 0.096 (b), and 0.092 (c)]. The red solid lines are the fitting curves using a two-gap model. The solid and dashed lines represent the $d$-wave and a single-gap BCS approach, respectively. The inset of (b) presents the temperature dependence of the magnetic penetration depths $\lambda _{ab}$.}
\label{PD}
\end{figure}

\begin{table*}
\caption{\label{tab:table 1}  {The superconducting transition temperature $T_{c}$ (in K), the SDW transition temperature $T_{N}$ (in K), the residual and normal-state electronic specific heat $\gamma_\textup{r}$ and $\gamma$$_{n}$, respectively (in mJ/mol K$^{2}$),  the universal parameter $\Delta C_\textup{el}/ \gamma_\textup{n} T_\textup{c}$, and the superconducting gap properties extracted from specific-heat and lower critical field ($H_{c1}$) measurements for BaFe$_{2-x}$Ni$_{x}$As$_{2}$ ($x$=0.10, 0.096, and 0.092) along with other 122 Fe-based superconductors.}}
\begin{ruledtabular}
\begin{tabular}{cccccccccccc}
Compounds &$T_c$  & $T_N$ & $\gamma_r$&$\gamma _{n}$ & $\Delta C_\textup{el}/ \gamma_\textup{n} T_\textup{c}$& $\Delta _{L}/k_{B}T_{c}$ & $\Delta _{S}/k_{B}T_{c}$ & $\Delta _{L}/\Delta _{S}$ & $\gamma_{1}, \gamma_{2}/\gamma_{n}$ &Technique & Ref.\\
\hline
BaFe$_{1.90}$Ni$_{0.10}$As$_{2}$ &20(1)&30(3)&1.6 &24.6 &1.04 & 1.85, 1.9 &0.79, 0.68 &2.3, 2.7&0.41, 0.59 & $C(T), \lambda _{ab}$&this work  \\

BaFe$_{1.904}$Ni$_{0.096}$As$_{2}$ &19(0.5)&32(5)&0.9 & 23.5& 1.06&1.8, 1.74& 0.74, 0.59 &2.4, 2.9&0.44, 0.56 & $C(T),\lambda _{ab}$&this work   \\

BaFe$_{1.908}$Ni$_{0.092}$As$_{2}$ &18.4(0.2)&39(4)& 2.6& 20.5& 1.12&1.74, 1.72& 0.68, 0.49 &2.5, 2.9 & 0.39, 0.61 & $C(T),\lambda _{ab}$&this work   \\

Ba(Fe$_{0.925}$Co$_{0.075}$)$_{2}$As$_{2}$&21.4&--&5.77 &23.8 & 1.2& 2.2&0.95 &2.3&0.33, 0.67 & $C(T)$&[\onlinecite{Hardy2010}] \\

Ba$_{0.6}$K$_{0.4}$Fe$_{2}$As$_{2}$& 35.8 &--& 1.2 & 50 & 1.54 & 2.88(0.2) & 0.64(0.02) & 4.45(0.3)&0.5, 0.5& $C(T),H _{c1}$&[\onlinecite{PP,CR}] \\

Ba$_{0.65}$Na$_{0.35}$Fe$_{2}$As$_{2}$ &29.4 &--&3.3 &57.5 & 1.26 &2.08& 1.06& 1.96 &0.48, 0.52 & $C(T)$&[\onlinecite{Pramanik2011}] \\
\end{tabular}
\end{ruledtabular}
\end{table*}

The temperature dependence of the $\lambda _{ab}$ applied along the $c$ axis is shown in the inset of Fig.\,5(b). At low temperatures from the inset, $\lambda _{ab}(T)$ does not show an exponential behavior as one would expect for a fully gapped clean $s$-wave superconductor. The main features in Fig.\,5, $\lambda(T)$-data, can be described in the following ways: (i) As the first step we compare our data to the $d$-wave and single-gap BCS theory under the weak-coupling approach (see solid and dashed lines in Fig.\,5). Indeed, both quantities lead to a rather different trend and show a systematic deviation from the data in the whole $T$-range below $T_{c}$. (ii) Then, the obtained temperature dependence of $\lambda^{-2}_{ab}(T)$ was analyzed by using the phenomenological $\alpha$-model. This model generalizes the temperature dependence of gap to allow $\alpha=2 \Delta(0)/T_c > 3.53$ ($i.e.$ $\alpha$ values higher than the BCS value). The temperature dependence of each energy gap for this model can be approximated as~\cite{Carrington}: $\Delta _{i}(T) = \Delta _{i}(0) {\tanh[1.82(1.018(\frac{T_{ci}}{T}-1))^{0.51}]}$, where $\Delta(0)$ is the maximum gap value at $T$ = 0. We adjust the temperature dependence of the London penetration depth by using the following expression:
\begin{equation}
 \frac{\lambda _{ab}^{-2}(T)}{\lambda _{ab}^{-2}(0)} = 1+\frac{1}{\pi}\int^{2 \pi}_0{ 2\int_{\Delta(T,\phi)}^{\infty}{\frac{\partial f}{\partial E} \frac{E dE d\phi}{\sqrt{E^2-\Delta^2(T,\phi)}}}},
\end{equation}
where $\Delta(T,\phi)$ is the order parameter as functions of temperature and angle. For the two-gap model, $\lambda^{-2}_{ab}$ is calculated as~\cite{Carrington}:
 {\begin{equation}
\lambda _{ab}^{-2}(T) = r\lambda _{1}^{-2}(T) + (1-r)\lambda _{2}^{-2}(T),
\end{equation}}where $0<r<1$. Equations (5) and (6) are used to introduce the two gaps and their appropriate weights.

The best description of the experimental data is obtained using values of $\Delta_{1}/k_{B}T_{c}$ = 1.72$\pm0.3$, 1.9$\pm0.3$ and 1.74$\pm0.25$, $\Delta_{2}/k_{B}T_{c}$ = 0.49$\pm0.3$, 0.68$\pm0.3$ and 0.59$\pm0.25$, and $r$ = 0.2$\pm0.1$, 0.32$\pm0.2$, and 0.48$\pm0.2$  for $x$ = 0.092, 0.096, and 0.1, respectively. The calculated penetration depth data are represented by the solid red lines in Fig.\,5. It is noteworthy that our extracted gap values fit to the two-band $s$-wave fit for Ba(Fe$_{1-x}$Co$_{x}$)$_{2}$As$_{2}$~\cite{FH}. Our investigated gap values for $H_{c1}$ and specific heat measurements have been found to be similar to those of values reported from the in-plane thermal conductivity~\cite{TH}. In addition, our results are consistent with the BaFe$_{2-x}$Co$_{x}$As$_{2}$ system, in which the superconducting energy gap does not contain a line of nodes anywhere on the Fermi surface, at any doping~\cite{CO}. On the other hand, the value of the gap amplitudes obtained for these SC samples scales relatively well with its $T_{c}$ in light of the recent results for the Fe-based superconductors~\cite{V12}. Interestingly, one can notice that the extracted ratio for the anisotropic $s$-wave order parameter $\alpha$ is smaller than the BCS value, which points to the existence of the large gap.

For the sake of comparison, we have summarized the $T_{c}$, the SDW transition $T_{N}$, $\gamma_\textup{r}$,  $\gamma$$_{n}$, the universal parameter $\Delta C_\textup{el}/ \gamma_\textup{n} T_\textup{c}$, and the values for the gaps $\Delta _{L}$, $\Delta _{S}$ for BaFe$_{1.908}$Ni$_{0.092}$As$_{2}$, BaFe$_{1.904}$Ni$_{0.096}$As$_{2}$, and BaFe$_{1.90}$Ni$_{0.10}$As$_{2}$ extracted from  specific-heat and lower critical field ($H_{c1}$) measurements along with other hole-doped 122 materials in Table I. The $\Delta _{L}$/$\Delta _{S}$ ratio of the investigated systems in this work is found to be lower than in Ba$_{0.6}$K$_{0.4}$Fe$_{2}$As$_{2}$~\cite{CR} and Ca$_{0.32}$Na$_{0.68}$Fe$_{2}$As$_{2}$~\cite{STJ} systems, but this ratio is higher than the Ba$_{0.65}$Na$_{0.35}$Fe$_{2}$As$_{2}$ sample extracted from earlier specific heat measurements (Table I). The gap magnitudes are scattered for different systems within the doped BaFe$_{2}$As$_{2}$. As mentioned above, the presence of a finite $\gamma_\textup{r}$ term is common in both electron- and hole-doped 122 compounds. Most remarkably, assuming a SC volume fraction in our investigated SC samples ($\gamma _{n} - \gamma _{r}$)/$\gamma _{n} \approx$ 87.3, 96.1, 93.4\% for $x$ = 0.092, 0.096, and 0.10, respectively, which is in fair agreement with our magnetization data. Additionally, the relative weight of each contributions illustrates that $\gamma_{2}/\gamma _{n}$ is always larger than $\gamma_{1}/\gamma _{n}$ indicating that the major gap develops around the Fermi surface sheet that exhibits the largest DOS. Theoretically, in a two-band model that $\frac{\gamma_{2}}{\gamma_{1}} \propto \sqrt{\frac{\Delta_{1}}{\Delta_{2}}}$ is expected in the interband weak-coupling limit~\cite{DOG}.

It is interesting to compare the present results with the other works for the most studied 122-based superconductors in which the electron pairing mechanisms are still fairly under debate. For hole-doped Ba$_{1-x}$K$_{x}$Fe$_{2}$As$_{2}$, heat-transport measurements have claimed the possibility of line nodes in the SC gap in the underdoped regime~\cite{Re1}. The similar nodal gap has also been observed in the heavily hole-overdoped Ba$_{0.1}$K$_{0.9}$Fe$_{2}$As$_{2}$ and KFe$_{2}$As$_{2}$~\cite{Ding3,M2,Re5}.  Interestingly, the isovalent substitution in Ba(Fe$_{0.64}$Ru$_{0.36}$)$_{2}$As$_{2}$ and BaFe$_{2}$(As$_{0.67}$P$_{0.33}$)$_{2}$ showed a large residual in thermal conductivity and $\sqrt{H}$ dependence, evidencing the presence of nodes in the SC gap~\cite{Re3}. For electron-doped systems similar to the current study, the field dependence of the specific heat of both underdoped and overdoped Ba(Fe$_{1-x}$Co$_{x}$)$_{2}$As$_{2}$ exhibits a Volovik-like nonlinear behavior, indicative of nodes in the SC gap~\cite{Re4}, while nodeless gaps have been reported in underdoped compounds~\cite{CO}. Penetration depth experiments with a careful analysis of the SC state on Ba(Fe$_{1-x}$Co$_{x}$)$_{2}$As$_{2}$ concluded the possibility of nodeless and nodes in the SC gap depending on the doping level~\cite{Re6,Re7}. Very recently, the SC gap structure of (Ba$_{1-x}$K$_{x}$)Fe$_{2}$As$_{2}$ was observed to vary with the composition from two nodeless isotropic SC gaps at the optimal doping to a strongly anisotropic gaps at the end of the SC dome at $x$ = 0.16~[\onlinecite{kim,Re8}]. In addition, the superfluid density of K$_{1-x}$Na$_{x}$Fe$_{2}$As$_{2}$ in the full temperature range follows a simple clean and dirty $d$-wave dependence, for pure and substituted samples, respectively~\cite{Re9}. Near optimal doping for both hole- and electron-doped 122 compounds, various experiments have clearly demonstrated multiple nodeless SC gaps~\cite{PRB,PP,FH,Ding2}. In fact, it is hard to get a simple pairing mechanism from such complex situation of the SC gap. The two independent techniques used here provide the self-consistent and convincing evidence for the nodeless gap in BaFe$_{2-x}$Ni$_{x}$As$_{2}$ superconductors covering the underdoped to the optimal doping.

Although in the current work we presented self-consistent data obtained from both magnetic penetration depth and specific heat measurements, some theoretical and other experiments also suggest a complicated pair symmetry for most iron-based superconductors, including various scenarios as mentioned above. However, it is important to emphasize that our investigated systems near optimal doping definitely underly and are consistent with nodeless multi-gaps in iron-arsenide multiband superconductivity in the presence of SDW, probably in the weak coupling regime.

\section{Conclusion}

To summarize, from an extensive thermodynamic study of high-quality BaFe$_{2-x}$Ni$_{x}$As$_{2}$ single crystals we have found that the magnetization loops exhibit a second peak, which is pronounced up to temperatures close to $T_{c}$. The main results are as follows. (i) Using the specific heat of a non-SC sample BaFe$_{1.75}$Ni$_{0.25}$As$_{2}$ as a reference, we are able to separate the electronic specific heat from the phonon contribution for the SC samples down to $T$ = 0.4\,K. (ii) Both the normal-state Sommerfeld coefficient and the jump of the specific heat $\Delta C/T_{c}$ are found to increase with Ni doping, indicating the strong competition between superconductivity and magnetism. (iii) For all our SC samples, the electronic specific heat displays a pronounced anomaly at $T_\textup{c}$ and a small residual part at low temperatures in the SC state. (iv) The observed temperature dependencies of $C_{el}/T$ and $\lambda^{-2}_{ab}$ are inconsistent with a single BCS gap as well as with a $d$-wave symmetry of the SC energy gap. Instead, our analysis is consistent with the presence of two $s$-wave-like gaps in the nearly optimally doped compounds.

\vspace{8mm}
\begin{acknowledgments}
We appreciate the useful discussions with Christoph Geibel, Robert Kuechler, Alexander Vasiliev, Helge Rosner, and R. Klingeler. Z.H. and J.Z. acknowledge the support from the Shanghai Pujiang Scholar program (No.13PJ1401100). The works in ECNU and IOP of CAS are supported by the Natural Science Foundation of China (Nos. 61125403, 11374011, and 91221303), the Ministry of Science and Technology of China (973 projects: Grants Nos. 2011CBA00110, 2012CB821400, 2013CB922301, and 2014CB921104), and The Strategic Priority Research Program (B) of CAS (Grant No. XDB07020300). The work at Rice is supported by National Science Foundation Grant No. DMR-1362219 and the Robert A. Welch Foundation Grant No. C-1839.
\end{acknowledgments}

\end{document}